\documentclass[12pt]{iopart}

\usepackage{iopams}
\usepackage{graphicx,amssymb}
\usepackage{color}
\newtheorem{lemma}{Lemma}
\newtheorem{proposition}{Proposition}

\newcommand{\be}{\begin{equation}}
\newcommand{\ee}{\end{equation}}
\def\dif{{\rm d}}

\begin{document}

\title[Thermodynamic class II Szekeres-Szafron singular models]
{Thermodynamic class II Szekeres-Szafron solutions. Singular models}

\author{Bartolom\'e Coll$^{1}$, Joan Josep Ferrando$^{1,2}$ and\\ Juan Antonio S\'aez$^3$}

\address{$^1$\ Departament d'Astronomia i Astrof\'{\i}sica, Universitat
de Val\`encia, E-46100 Burjassot, Val\`encia, Spain}

\address{$^2$\ Observatori Astron\`omic, Universitat
de Val\`encia, E-46980 Paterna, Val\`encia, Spain}

\address{$^3$\ Departament de Matem\`atiques per a l'Economia i l'Empresa,
Universitat de Val\`encia, E-46022 Val\`encia, Spain}

\ead{bartolome.coll@uv.es; joan.ferrando@uv.es; juan.a.saez@uv.es}

\begin{abstract}
A family of parabolic Szekeres-Szafron class II solutions in local thermal equilibrium is studied and their associated thermodynamics are obtained. The subfamily with the hydrodynamic behavior of a generic ideal gas (defined by the equation of state $p = k n \Theta$) results to be an inhomogeneous generalization of flat FLRW $\gamma$-law models. Three significative interpretations that follow on from the choice of three specific thermodynamic schemes are analyzed in depth. First, the generic ideal gas in local thermal equilibrium; this interpretation leads to an inhomogeneous temperature $\Theta$. Second, the thermodynamics with homogeneous temperature considered by Lima and Tiomno (CQG {\bf 6} 1989). And third, a new model having exactly the homogeneous temperature of the FLRW limit. It is shown that the three models above fulfill the necessary macroscopic requirements for physical reality (positivity of matter density and temperature, energy conditions and compressibility conditions) in wide domains of the spacetime. 
\end{abstract}
%

%


\section{Introduction}
\label{sec-intro}

Szekeres cosmological models \cite{Szekeres} are dust inhomogeneous perfect fluid solutions that can describe the Universe in the post-recombination era \cite{Krasinski,Krasinski-Plebanski,Krasinski-et-all}. These models were generalized by Szafron \cite{Szafron} by considering a non-vanishing pressure \cite{Krasinski, Krasinski-Plebanski}. Several papers have been devoted to perform an invariant characterization of the Szekeres-Szafron metrics \cite{Wainwright, Szafron-Collins, Barnes-Row} (see also \cite{Krasinski,Krasinski-Plebanski}), and we have recently achieved an IDEAL approach to these solutions \cite{FS-SS}. 

The physical and geometric properties of the pioneer dust solutions by Szekeres and of the Szafron models with constant pressure have been widely analyzed in the literature \cite{Krasinski, Krasinski-Plebanski, bonnor, Bonnor-ST, berger, goode} (see also the recent papers \cite{hellaby, G-hellaby} and references therein). Nevertheless, the physical meaning of the full set of Szekeres-Szafron (SS) metrics is still an open problem. Thus, some authors have remarked on the difficulties in associating a realistic equation of state to these solutions \cite{Krasinski-et-all} \cite{Lima-Tiomno-a}. However, a few isolated results can be quoted. 

Lima and Tiomno \cite{Lima-Tiomno-a} considered a subfamily of class II SS metrics that extend the Szekeres subfamily considered by Bonnor and Tomimura \cite{Bonnor-T} to non-vanishing pressure. These models evolve to a FLRW era and have been proposed as two-fluid cosmologies \cite{Lima-Tiomno-a}. In a subsequent paper \cite{Lima-Tiomno-b} a subset of the parabolic solutions has been interpreted as one-component thermodynamic fluids. A wider family with a similar thermodynamic scheme was considered in \cite{QS-1995}.

Krasi\'nski {\em et al.} \cite{KQS} proved that there are thermodynamic Szekeres-Szafron solutions of class II without symmetries. Nevertheless, if a class I Szekeres-Szafron metric admits a thermodynamic scheme then, necessarily, it admits symmetries \cite{KQS}. The latter result has recently been recovered in \cite{FS-SS}. 

We have set ourselves the goal of studying in detail the SS spacetimes that model the evolution of a thermodynamic perfect fluid in local thermal equilibrium, and to analyze thermodynamics that fulfill the necessary macroscopic constraints for physical reality. A fundamental tool in this study is the hydrodynamic approach to the concept of local thermal equilibrium developed in \cite{Coll-Ferrando-termo} and \cite{CFS-LTE}, and the necessary constraints for physical reality analyzed in \cite{CFS-CC} from a hydrodynamic point of view. This procedure has being applied in a recent paper \cite{CFS-CIG} to analyze classical ideal gas solutions.

It is worth remarking that our study is based in a macroscopic approach without any reference to statistical mechanics or kinetic theory. It follows the line of research of the above quoted papers \cite{Coll-Ferrando-termo, CFS-LTE, CFS-CC, CFS-CIG}, and it is supported by historical and recent references on macroscopic relativistic thermodynamics \cite{Tolman-30, Tolman, Eckart, Taub, Israel, Plebanski, Lichnero-1, Israel-b, Israel-St, Anile, Lichnero-2, Jou-Casas, ReZa}.

In section \ref{sec-classII-termo} we analyze the local thermal equilibrium condition for the SS metrics of class II, and we show that three families arise in a natural way: the {\em singular models} (parabolic models with a linear dipole term), the {\em regular models} and the metrics admitting a three-dimensional isometry group $G_3$ on space-like two-dimensional orbits $S_2$.

In this paper we focus on the singular models. In section \ref{sec-parabolic-0} we obtain the canonical form for the metric line element and we offer the expression for the hydrodynamic quantities: {\em energy density} $\rho$ and {\em pressure} $p$. We also acquire their associated thermodynamics, namely, we determine the {\em specific entropy} $s$, the {\em matter density} $n$ and the {\em temperature} $\Theta$. Finally, we obtain an implicit expression for the {\em indicatrix of the local thermal equilibrium}, $\chi = \chi (\rho,p)$, which gives the square of the speed of sound in terms of the hydrodynamic quantities, $c_s^2 = \chi(\rho, p)$. 

The equation of state $p = k n \Theta$ does not determine the characteristic equation of a fluid, and we must add a relation between the specific internal energy and the temperature, $\epsilon = \epsilon(\Theta)$, in order to obtain all the thermodynamic properties of the fluid. For example, when $\epsilon = c_v \Theta$, we have a {\it classical ideal gas}. If we do not fix $\epsilon(\Theta)$, we have a set of fluids submitted to the equation of state $p = k n \Theta$ which we name {\it generic ideal gases}. In  \cite{CFS-LTE} we have shown that the indicatrix function of the generic ideal gases is of the form $\chi = \chi(\pi)$, $\pi \equiv \rho/p$. Moreover, the study of the named inverse problem \cite{CFS-LTE} shows that this hydrodynamic property can be fulfilled by fluids other than the generic ideal gas. Section \ref{sec-chi-pi} is devoted to determining the singular models which are compatible with an indicatrix function of the form $\chi = \chi(\pi)$. We have that these {\em ideal singular models} have the hydrodynamic behavior of a generic ideal gas and evolve to the FLRW $\gamma$-law models. We study the necessary macroscopic conditions for physical reality of the solutions by analyzing the energy condition \cite{Plebanski} and the relativistic compressibility conditions \cite{CFS-CC, Israel, Lichnero-1}. Their associated thermodynamics are also outlined.

In the following sections we analyze in detail three of the all possible thermodynamic schemes that can be associated with the ideal singular models. And we show that they fulfill suitable requirements for physical reality in a wide domain of the spacetime. In section \ref{sec-idelagas} we consider a generic ideal gas scheme and obtain all the thermodynamic quantities by using the results in \cite{CFS-LTE}. In section \ref{sec-LT} we revisit the thermodynamic scheme of the model presented by Lima and Tiomno \cite{Lima-Tiomno-b}, which has homogeneous temperature and is thus compatible with non-vanishing conductivity. And in section \ref{sec-T-FLRW} we present a new thermodynamic scheme, which has the same (homogeneous) temperature as the $\gamma$-law models of the FLRW limit.

Finally, in section \ref{sec-summary} we present a discussion of the results and some tables that summarize the main characteristics of the models.


\section{Thermodynamic Szekeres-Szafron metrics of class II}
\label{sec-classII-termo}

The canonical form of the Szekeres-Szafron metrics of classes I and II can be
found in several papers \cite{Krasinski, Krasinski-Plebanski, Szafron}.  With a slightly different notation, the metric line element of class II solutions takes the expression:
\begin{equation} \label{SS-canonica}
\dif s^2 = - \dif t^2 + \phi^2[(B + P)^2 \dif z^2 + C^2 (\dif x^2 + \dif y^2)] \, ,
\end{equation}
where 
\begin{eqnarray} \label{SS-metricfunctions}
\phi = \phi(t)  \,  , \qquad B=B(t,z) \,  , \qquad P= S \,C \, , \\[1mm]
C = C(x,y) \equiv [1 + \frac{k}{4} (x^2 + y^2)]^{-1}, \qquad k \equiv 0, 1, -1  , \label{II-C} \\[1mm] 
S = S(z,x,y) \equiv  \frac12 U(z) (x^2 + y^2) + V_1 (z) x +  V_2 (z) y + 2\,W(z) \,   .  \label{II-S}  
\end{eqnarray}
These metrics are perfect fluid solutions when the above metric functions fulfill:
\begin{equation} 
\ddot{B} + \frac{3 \, \dot{\phi}}{\phi} \dot{B} -  \frac{k}{\phi^2} B = \frac{1}{\phi^2}(U + kW) \, . \label{eq-SS-II}
\end{equation}
Moreover, the pressure $p$ and the energy density $\rho$ are given by: 
\begin{eqnarray} \label{pressure-II}
p = -\left[\frac{2\, \ddot{\phi}}{\phi} + \frac{\dot{\phi}^2}{\phi^2} + \frac{k}{\phi^2}\right]   \,  ,  \\[1mm]
\rho = \frac{3 \, \dot{\phi^2}}{ \phi^2} + \frac{3k}{\phi^2} + \frac{2 \, [\phi \, \dot{\phi}\, \dot{B} - k(B + W) -U]}{\phi^2 (B + P)}   \,  ,
\label{density-II}
\end{eqnarray}
and the unit velocity $u$ is geodesic and its expansion is:
\begin{equation} \label{expansion-II}
\theta = \frac{3 \, \dot{\phi}}{ \phi} + \frac{\dot{B}}{B + P}  \,  .
\end{equation}
The spacetime is Petrov-Bel type D and the simple Weyl eigenvalue is:
\begin{equation} \label{omega-II}
\omega = - \frac{\phi \, \dot{\phi} \dot{B}- k(B + W) - U }{3 \phi^2(B + P)} = \frac12 \left[\frac{\dot{\phi^2}}{ \phi^2} + \frac{k}{\phi^2}\right] - \frac{\rho}{6}  \,  .
\end{equation}

Under the hypothesis $\rho + p \not=0$, the FLRW limit follows in the conformally flat case ($\omega=0$). In terms of the metric functions this condition holds if, and only if, $\dot{B}=0$ (and then $k(B+W)+U=0$). Moreover, $P$ can be redefined so that we can take $B=0$. Hereinafter we consider {\em strict Szekeres-Szafron metrics} of class II, that is, metrics of the form (\ref{SS-canonica}) with $\dot{B}\not=0$.

The strict SS metric (\ref{SS-canonica}) admits a $G_3$ on $S_2$ if, and only if, $\rho  = \rho(t,z)$, that is $P= P(z)$ or, equivalently, $V_1 = V_2 =0$ and $U = k W$ \cite{FS-SS}. In this case, we can redefine the function $B$ so that $P=0$. 

Finally, the non-conformally flat barotropic limit follows when, in addition to having a $G_3$ on $S_2$, the function $B+P$ factorizes. Now we can redefine function $B$ and coordinate $z$ so that $P=0$ and $B=B(t)$, and we obtain the canonical form of the Kantowski-Sachs metrics  and of their parabolic and hyperbolic counterparts \cite{Krasinski, K-S}.


\subsection{Thermodynamic constraints for class II Szekeres-Szafron metrics}

A relevant question in studying perfect fluid solutions is to analyze their interpretation as reasonable physical media. Pleba\'nski \cite{Plebanski} energy conditions are necessary algebraic conditions for physical reality and, in the perfect fluid case, they state: $-\rho < p \leq \rho$. The determination of the spacetime regions where these constraints hold is a basic query in analyzing a given perfect fluid solution. 

Furthermore, if we want the solution to describe a thermodynamic perfect fluid in local thermal equilibrium we must impose complementary restrictions. From a macroscopic point of view, a necessary condition for the fluid to admit a thermodynamics is that a function $n$ exists such that \cite{CFS-LTE}:
\begin{equation} \label{lte-n}
\dot{n} + n \theta =0 \, , \qquad   \dif n \wedge \dif p \wedge \dif \rho = 0 \, .
\end{equation}
Then, the function of state $n=n(\rho,p)$ is the conserved {\em matter density} of the fluid. Moreover we can identify the (absolute) {\em temperature} $\Theta$ of the fluid and the {\em specific entropy} $s$ as the functions submitted to the {\em local thermal equilibrium equation}:
\begin{equation} 
\Theta \dif s = (1/n) \dif \rho + (\rho+p) \dif (1/n) \, ,   \label{re-termo}
\end{equation}
and the {\em specific internal energy} $\epsilon$ is defined by the relation:
\be \label{epsilon}
\rho= n(1+\epsilon) \, .
\ee

We have already shown \cite{Coll-Ferrando-termo}  \cite{CFS-LTE} that the macroscopic notion of local thermal equilibrium admits a purely hydrodynamic formulation: {\em a non isoenergetic ($\dot{\rho} \not= 0$) perfect energy tensor $T$ evolves in local thermal equilibrium if, and only if, the hydrodynamic quantities $(u, \rho, p)$ fulfill the hydrodynamic sonic condition}: 
\begin{equation} \label{lte-chi}
\   \dif \chi \wedge \dif p \wedge \dif \rho = 0 \, , \qquad \chi \equiv \frac{\dot{p}}{\dot{\rho}}   \, .
\end{equation}
Then, the {\em indicatrix of the local thermal equilibrium} $\chi$ is a function of state, $\chi = \chi(\rho,p)$, which physically represents the square of the {\em speed of sound} in the fluid, $\chi (\rho ,p) \equiv  c^2_{s}$.

Our first goal in this paper is to analyze the local thermal equilibrium condition for class II Szekeres-Szafron metrics that are expanding $\theta \not=0$ and non barotropic, $\dif \rho \wedge \dif p \not=0$. We have then $\dot{\rho} \not=0$ and $\dot{p} \not=0$. For the SS metrics the hydrodynamic sonic condition (\ref{lte-chi}) admits an equivalent and simpler expression \cite{FS-SS}:
\begin{equation} \label{lte-theta}
\   \dif \theta \wedge \dif \rho \wedge \dif t = 0 \,  .
\end{equation}

For expanding and non barotropic SS metrics of class II given in (\ref{SS-canonica}), we can substitute in (\ref{lte-theta}) the expressions (\ref{density-II}) of the energy density and (\ref{expansion-II}) of the expansion. Then, we obtain:
\begin{equation} \label{lte-II}
\   \dif \! \left[\frac{\dot{B}}{B+P}\right]  \wedge \dif \! \left[\frac{k(B+W)+U}{B+P}\right]  \wedge d t = 0 \,  .
\end{equation}
If $k=0$ and $U=0$ ({\em singular models}) the local thermal equilibrium condition (\ref{lte-II}) identically holds. Otherwise, if $k^2 + U^2 \not =0$ ({\em regular models}), condition (\ref{lte-II}) is equivalent to:
\begin{equation} \label{lte-II-b}
\   \dif \! \left[\frac{\dot{B}}{k(B+W)+U}\right]  \wedge \dif  P \wedge \dif t = 0 \,  .
\end{equation}
When the metric admits a $G_3$ on $S_2$ we have $P=P(z)$ and (\ref{lte-II-b}) identically holds, according to the well-known result that a perfect fluid solution with these symmetries always admits a thermodynamic scheme. Otherwise we have $P_{\!x}^{\, 2} + P_{\!y}^{\, 2} \not=0$, and (\ref{lte-II-b}) becomes equivalent to:
\begin{equation} \label{lte-II-c}
\left[\frac{\dot{B}}{k(B+W)+U}\right]' = 0 \,  .
\end{equation}
where, for any function $q$, $q' = \partial_z q$. Thus, we have shown:
\begin{proposition} \label{prop-II}
An expanding and non barotropic Szekeres-Szafron metric of class II is a thermodynamic perfect fluid solution in local thermal equilibrium if, and only if, it satisfies at least one of the following three conditions:
\begin{itemize}
\item[(i)] 
It admits a $G_3$ on $S_2$, that is, $P=P(z)$.
\item[(ii)] 
It is a singular model, that is, $k = U =0$.
\item[(iii)] 
It is a regular model, that is, $k^2 + U^2 \not =0$ and condition {\em (\ref{lte-II-c})} holds.
\end{itemize}
\end{proposition}
Note that these three conditions do not determine a classification of the expanding and non barotropic class II SS metrics in local thermal equilibrium. Indeed, the singular models defined in (ii) are compatible with the condition $P=P(z)$ of (i), and thus they contain a subfamily of metrics admitting a $G_3$ on $S_2$. Moreover $P=P(z)$ does not imply condition (\ref{lte-II-c}) and the three subfamilies are necessary to cover all the class II SS metrics that are expanding and non barotropic.

Consequently, the full study of the thermodynamic class II SS solutions involves analyzing the three cases considered in proposition \ref{prop-II}. The main aim 
of this paper is to study in depth the singular models by obtaining their associated thermodynamics and by outlining some physically relevant models. The study of the other two cases is a work in progress which will be presented elsewhere. Some preliminary results for case (iii) were reported years ago in \cite{cfERE}. It is also worth mentioning the paper by Krasi\'nski {\em et al.} \cite{KQS} where, by starting from condition  (\ref{lte-n}), they proved the existence of thermodynamic class II Szekeres-Szafron solutions without symmetries.


\section{Singular models}
\label{sec-parabolic-0}


\subsection{Metric and hydrodynamic quantities: energy density and pressure} 
\label{subsec-metric-parabolic}

For a singular model ($k = U =0$) the field equation (\ref{eq-SS-II}) becomes $\ddot{B} + \frac{3 \, \dot{\phi}}{\phi} \dot{B} =0$, a linear equation whose general solution is of the form $B(t,z)=\alpha(t) c(z) + b(z)$, where $a(z)$ and $b(z)$ are arbitrary real functions, and $\alpha(t)$ is a particular solution to the equation:
\be  \label{eq-alpha-b}
\dot{\alpha} = - \frac{1}{\phi^3}  \, .
\ee
On the other hand, if we write $c(z) = \varepsilon f(z)$, $\varepsilon = 0, 1$, and $Q = (b + P)/f$, and we change the coordinate $z$ as $d \tilde{z} = f(z) d z$, we obtain the following {\em canonical form of the singular models}:
\begin{equation} \label{parabolic-SS-canonica}
\dif s^2 = - \dif t^2 + \phi^2[(\varepsilon \alpha + Q)^2 \dif z^2 + \dif x^2 + \dif y^2] \, ,
\end{equation}
where $\phi = \phi(t)$ and $\alpha=\alpha(t)$ are constrained by the equation (\ref{eq-alpha-b}), and
\begin{equation} \label{parabolic-SS-metricfunctions}
Q =  V_1 (z) x +  V_2 (z) y + 2\,W(z) \,   .   
\end{equation}
The {\em pressure} $p$ and the {\em energy density} $\rho$ are then given by: 
\begin{eqnarray} \label{pressure-parabolic}
p = -\left[\frac{2\, \ddot{\phi}}{\phi} + \frac{\dot{\phi}^2}{\phi^2}\right]   \,  ,  \\[1mm]
\rho = \frac{3 \, \dot{\phi^2}}{ \phi^2} - \frac{2 \, \varepsilon \, \dot{\phi}}{\phi^4 (\varepsilon \alpha + Q)}   \,  .
\label{density-parabolic}
\end{eqnarray}
And the {\em expansion} of the fluid is:
\begin{equation} \label{expansion-parabolic}
\theta = \frac{3 \, \dot{\phi}}{ \phi} - \frac{\varepsilon}{\phi^3(\varepsilon \alpha + Q)} = \partial_t [\ln\{\phi^3( \varepsilon \alpha + Q)\}] \,  .
\end{equation}
The singular SS metrics of class II (\ref{parabolic-SS-canonica}) depend on an arbitrary function of time ($\phi(t)$ and $\alpha(t)$ are submitted to constraint (\ref{eq-alpha-b})) and three arbitrary functions, $V_1 (z)$, $V_2 (z)$ and $W(z)$, of the coordinate $z$. Now, we recover the FLRW limit by making $\varepsilon=0$, and the barotropic limit follows if $Q=constant$. The metric admits a $G_3$ on flat two-dimensional orbits when $V_1 = V_2 =0$. 
%


\subsection{Thermodynamic scheme: entropy, matter density and temperature} 
\label{subsec-scheme-parabolic}

We know that the singular SS metrics of class II (\ref{parabolic-SS-canonica}) define perfect fluid solutions in l.t.e., and the hydrodynamic quantities pressure and energy density are given in (\ref{pressure-parabolic}) and (\ref{density-parabolic}), respectively. Now we shall solve what we have termed the inverse problem \cite{CFS-LTE} for these solutions, namely, we shall obtain the full set of associated thermodynamic quantities: specific entropy $s$, matter density $n$ and temperature $\Theta$.

For a conservative perfect energy tensor $T$ in l.t.e. the range of associated thermodynamics depends on two arbitrary functions of the specific entropy. More precisely \cite{CFS-LTE}, each thermodynamics is determined by a specific entropy $s$ and a matter density $n$ of the form $s=s(\bar{s})$, $n=\bar{n} N(\bar{s})$, where $s(\bar{s})$ and $N(\bar{s})$ are arbitrary functions, and where $\bar{s} = \bar{s}(\rho,p)$ is a particular solution to the local adiabatic condition $\dot{s}=0$ and $\bar{n} =\bar{n}(\rho,p)$ is a particular solution to the matter conservation equation (\ref{lte-n}). 

For the sake of clarity, from now on we write the hydrodynamic quantities in terms of the {\em Hubble function} $H$:
\begin{eqnarray} \label{pressure-parabolic-H}
p = -{2\, \dot{H}} - 3 H^2    \,  , \qquad   H \equiv \frac{\dot{\phi}}{\phi} \, , \\[1mm]
\rho = 3 H^2 - \frac{2 \, \varepsilon \, H}{\phi^3 (\varepsilon \alpha + Q)}   \,  .
\label{density-parabolic-H}
\end{eqnarray}
For a strict SS metric ($\varepsilon= 1$), we can isolate the function $Q$ from (\ref{density-parabolic-H}), and we obtain:
\be \label{Q-rho-p}
Q =  \left[\frac{2 H}{\phi^3 (3 H^2 - \rho)} - \alpha \right] \equiv Q(\rho,p) \, .
\ee
Note that $\phi, \alpha$ and $H$ are functions of $p$ because $\dot{p}\not=0$ and then $t=t(p)$. Therefore $Q$ is a function of state, $Q= Q(\rho,p)$. Moreover $\dot{Q}=0$ as a consequence of (\ref{parabolic-SS-metricfunctions}), and then we can take $\bar{s}= Q$.

On the other hand, from expression (\ref{expansion-parabolic}) of the expansion it follows that $\bar{n} = [\phi^3 (\varepsilon \alpha + Q)]^{-1}$ is a solution to equation (\ref{lte-n}). Thus we have shown:
\begin{proposition} \label{prop-s-n-parabolic}
The thermodynamic schemes associated with the singular models {\em (\ref{parabolic-SS-canonica})} are determined by a specific entropy $s$ and a matter density $n$ of the form:
\be  \label{s-n-parabolic}
s = s(Q) \equiv s(\rho, p) \, , \qquad \quad n = \frac{N(Q)}{\phi^3 (\varepsilon \alpha + Q)} \equiv n(\rho,p) \, .  
\ee
\end{proposition}
Note that, $Q$ being a function of state as a consequence of (\ref{Q-rho-p}), then $s$ and $n$ are too: $s=s(\rho, p)$, $n=n(\rho,p)$. 

In order to determine the temperature associated with each one of the thermodynamic schemes defined by the pair $\{s,n\}$ given in (\ref{s-n-parabolic}) we can start from the local thermal equilibrium equation (\ref{re-termo}) that can be written as:
\begin{equation} 
\Theta \dif s = \dif \, h - \frac{1}{n} \dif \, p \, , \qquad \quad  h \equiv 1 + \epsilon + \frac{p}{n}  = \frac{\rho +p}{n} \, ,  \label{re-termo-f}
\end{equation}
where $h$ is the {\em relativistic specific enthalpy}.

For the singular models the specific enthalpy can be calculated from (\ref{pressure-parabolic-H}), (\ref{density-parabolic-H}) and (\ref{s-n-parabolic}), and we obtain ($\varepsilon=1$):
\begin{equation} 
h = \frac{\rho +p}{n} = r(Q) [\lambda(t) + \mu(t) Q] \, , \qquad r(Q) = [N(Q)]^{-1}  \, ,  \label{index-f}
\end{equation}
\be \label{lambda-mu}
\lambda(t) \equiv - 2 [H + \phi^3 \alpha \dot{H}] \, , \qquad  \mu(t) \equiv -2 \phi^3 \dot{H}  \, .
\ee
Then, from (\ref{re-termo-f}) and (\ref{index-f}) we have:
\be  \label{temperatura-par}
\Theta = \left(\frac{\partial h}{\partial s}\right)_p = \frac{1}{s'(Q)} \left(\frac{\partial h}{\partial Q}\right)_t = \frac{1}{s'} [r' \lambda + (r' Q + r)\mu]   \, .
\ee
Consequently, we can state:
\begin{proposition} \label{prop-T-parabolic}
For the singular models {\em (\ref{parabolic-SS-canonica})}, the temperature of the thermodynamic schemes given in proposition {\em \ref{prop-s-n-parabolic}} takes the expression:
\be  \label{T-parabolic}
\Theta = \ell(Q) \lambda(t) + m(Q) \mu(t)  \, , 
\ee
where $\lambda(t)$ and $ \mu(t)$ are given in {\em (\ref{lambda-mu})} and
\be  \label{T-parabolic-b}
 \ell(Q) \equiv \frac{r'}{s'} \, , \quad  m(Q) \equiv \frac{1}{s'}[Q r' + r]  \, , \quad r(Q) \equiv \frac{1}{N}  \, .
\ee
\end{proposition}
%
 

\subsection{The indicatrix function: speed of sound}
\label{subsec-chi-parabolic}

With the aim of increasing our knowledge of the physical qualities of the models it is convenient to know the  function of state that gives the square of the speed of sound in terms of the hydrodynamic quantities, $c_s^2 = \chi(\rho,p)$. When the hydrodynamic sonic condition (\ref{lte-chi}) holds, this function is equal to the indicatrix of the l.t.e., $\chi = \dot{\rho}/\dot{p}$, that is, it can be evaluated from the hydrodynamic quantities $(u, \rho, p)$. Thus, it is independent of the thermodynamic schemes $\{s,n\}$ defined by the two arbitrary functions $s(Q)$ and $r(Q)$, and it only provides information on the hydrodynamic properties of the fluid.

As we will see in the following sections the interest in obtaining $\chi(\rho,p)$ is twofold. On one hand, we can impose on it complementary conditions that define a specific family of fluids; for example, in the case of a generic ideal gas we have $\chi = \chi(\pi)$, $\pi= \rho/p$ \cite{CFS-LTE}. On the other hand, it is a useful tool to impose the relativistic compressibility conditions and thus to enable good physical behavior of the models \cite{CFS-CC}. 

From (\ref{density-parabolic-H}) we can compute $\dot{\rho}$ by taking into account (\ref{eq-alpha-b}) and (\ref{Q-rho-p}), and we obtain:
\begin{eqnarray} \label{rho-punt}
\dot{\rho} = A \rho^2 + B \rho + C \, , \\ A \equiv - \frac{1}{2H} \, , \quad  B \equiv  \frac{\dot{H}}{H} \, , \quad  C \equiv 3 H(\dot{H} + \frac{3}{2} H^2) = - \frac32 H p \, . \label{rho-punt-b}
\end{eqnarray}
Consequently, we have the following result:
\begin{proposition} \label{prop-chi-parabolic}
For the singular models {\em (\ref{parabolic-SS-canonica})}, the square of the speed of sound takes the expression:
\be  \label{chi-parabolic}
c_s^2  = \chi(\rho,p)  \equiv \frac{1}{{\cal A}(p) \rho^2 + {\cal B}(p) \rho + {\cal C}(p)} \, ,
\ee
where ${\cal A}$, ${\cal B}$ and ${\cal C}$ are the functions of $t$ (and then of $p$) given by:
\begin{equation} 
{\cal A}(p) \equiv - \frac{1}{2 H \dot{p}} \, , \qquad {\cal B}(p) \equiv  \frac{\dot{H}}{ H \dot{p}} \, ,  \qquad {\cal C}(p) \equiv  - \frac{3 H p}{2 \dot{p}}  \label{ABCcal} \, .   
\end{equation}
\end{proposition}
Note that (\ref{chi-parabolic}) provides an expression of the indicatrix function which is implicit in the variable $p$. For a specific choice of the metric function $\phi(t)$ we can obtain $p(t)$ from (\ref{pressure-parabolic-H}), and then we can get $t(p)$. Then, the explicit form of $\chi(\rho,p)$ can be obtained (see forthcoming sections).

If we impose the compressibility conditions on this implicit generic expression of $\chi(\rho,p)$  we would get inequalities involving third order derivatives of the metric functions. This is not difficult to do but the result would not have a practical application. In the following sections we analyze the compressibility conditions for a specific family of solutions.


\section{Models with the hydrodynamic behavior of a generic ideal gas}
\label{sec-chi-pi}


\subsection{Metric and hydrodynamic quantities: energy density, pressure and speed of sound} 
\label{subsec-metric-parabolic-ideal}

Now we analyze when the singular models considered above are compatible with the equation of state of a {\em generic ideal gas}, namely:
\begin{equation}
p = k n \Theta  \, , \qquad \quad    k \equiv {k_B \over m} \,  .  \label{gas-ideal}
\end{equation}
In \cite{CFS-LTE} we have solved the direct problem for the generic ideal gases by studying the hydrodynamic constraints that equation (\ref{gas-ideal}) imposes, and we have shown:

\begin{lemma}
A perfect energy tensor $T=(u,\rho,p)$ represents the evolution of a generic ideal gas in l.t.e. if, and only if, the indicatrix function is of the form:
\be \label{chi-gas-ideal}
\chi = \chi(\pi) \not= \pi \, , \qquad \chi = \frac{\dot{p}}{\dot{\rho}} \, , \qquad \pi = \frac{p}{\rho} \, .
\ee
\end{lemma}

From the expression of the indicatrix function (\ref{chi-parabolic}) we obtain that, for the singular models, the generic ideal gas constraint (\ref{chi-gas-ideal}) is equivalent to:
\be
{\cal A} p^2 = c_1 \, \qquad {\cal B} p = c_2 \, \qquad {\cal C} = c_3 \, , \qquad c_i = constant \, .
\ee
Then, if we use the expressions (\ref{ABCcal}) and (\ref{pressure-parabolic-H}), and we take into account that $\dot{p}\not=0$, we obtain:
\begin{lemma}
The singular models with an indicatrix function of the form {\em (\ref{chi-gas-ideal})} fulfill the equations: 
\begin{equation} \label{eq-parabolic-gas-ideal}
\hspace{-10mm} \frac{1}{H} = -2 c_1 \frac{\dot{p}}{p^2}  \, , \qquad \frac{\dot{H}}{H} = c_2 \frac{\dot{p}}{p} \, , \qquad 3H = - 2c_3 \frac{\dot{p}}{p}  \, , \qquad p = -{2\, \dot{H}} - 3 H^2  \, ,
\end{equation}
where $c_i$ are non-vanishing constants.
\end{lemma}
We have $H=\dot{\phi}/\phi$. Thus, conditions (\ref{eq-parabolic-gas-ideal}) constitute a third-order differential system for the metric function $\phi(t)$. We can integrate the third equation and obtain:
\be \label{p-phi-0}
p = C \phi^{-3 \gamma} \, \qquad C = constant \, , \qquad \gamma \equiv \frac{1}{2c_3} \, .
\ee
And from the second equation in (\ref{eq-parabolic-gas-ideal}) we get $\dot{\phi}  = \kappa \, \phi^{1-3 \gamma c_2}$. Then, we can obtain $H$, $\dot{H}$ and $\dot{p}$ as a power function of $\phi$. The resulting expressions are compatible with equations (\ref{eq-parabolic-gas-ideal}) if, and only if, 
\be  \label{ki-gamma}
c_1 = \frac{\gamma - 1}{2 \gamma} \, , \qquad c_2 = \frac{1}{2} \, , \qquad  c_3 = \frac{1}{2 \gamma} \, , \qquad C = 3 \kappa^2 (\gamma - 1)   \, .
\ee
Consequently, 
\be  \label{phi-punt}
\dot{\phi}  = \kappa \, \phi^{1-3 \gamma/2} \,  \, .
\ee

On the other hand, from (\ref{eq-alpha-b}) and (\ref{phi-punt}) we obtain $\alpha'(\phi) = \dot{\alpha}/{\dot{\phi}} = \kappa^{-1} \phi^{\frac32 \gamma -4}$, and then:
\be \label{alpha-0}
\alpha = \alpha(\phi) \equiv
\cases{
\alpha_0 + \frac{2 \, \phi^{\frac32 (\gamma -2)}}{3 \kappa (2 - \gamma)}  \, , \qquad {\rm if} \quad \gamma \not= 2  \cr
\alpha_0 - \frac{1}{\kappa} \,  \ln \phi \, , \qquad \quad  \ \    {\rm if} \quad  \gamma = 2
}
\ee
Note that the function $\alpha$ appears in the metric expression (\ref{parabolic-SS-canonica}) through $\alpha + Q$. Thus we can redefine $W(z)$ such that we can take $\alpha_0 = 0$. Moreover, we can integrate equation (\ref{phi-punt}), and considering (\ref{density-parabolic}), (\ref{p-phi-0}), (\ref{ki-gamma}) and (\ref{alpha-0}), we arrive to:
\begin{proposition} \label{prop-parabolic-ideal}
The singular models with an indicatrix function of generic ideal gas type have a metric line element of the form {\em (\ref{parabolic-SS-canonica})}, where $Q$ is specified in {\em (\ref{parabolic-SS-metricfunctions})} and $\phi(t)$ and $\alpha(t)$ are given, respectively, by:
\begin{equation} \label{phi-t}
\phi(t) = \left[\frac32 \kappa \gamma \, t + \bar{\phi}_0 \right]^{\frac{2}{3 \gamma}}   ,
\end{equation}
\be \label{alpha-1}
\alpha(t) = \alpha(\phi) \equiv
\cases{
 \frac{2 \, \phi^{\frac32 (\gamma -2)}}{3 \kappa (2 - \gamma)}  , \qquad {\rm if} \quad \gamma \not= 2  \cr
- \frac{1}{\kappa} \,  \ln \phi  , \quad \quad \,  \  \   {\rm if} \quad  \gamma = 2
}
\ee
Moreover the pressure $p$ and the energy density $\rho$ are:
\begin{eqnarray} \label{pressure-parabolic-ideal}
p = \frac{3 \kappa^2 (\gamma-1) }{\phi^{3 \gamma}}    \,  ,  \\[1mm]
\rho = \frac{3 \kappa^2 }{\phi^{3 \gamma}} - \frac{2 \, \varepsilon \, \kappa}{\phi^{3(1 + \frac{\gamma}{2})} (\varepsilon \alpha + Q)} \, .
\label{density-parabolic-ideal}
\end{eqnarray}
And the speed of sound is given by:
\be \label{chi-parabolic-ideal}
c_s^2  = \chi(\pi)  \equiv \frac{2 \, \gamma \, \pi^2}{ (\pi + 1) (\pi + \gamma-1)} \, , \qquad \pi \equiv \frac{p}{\rho} \, .
\ee
\end{proposition}
%


\subsection{Analysis of the solutions. Energy conditions} 
\label{subsec-energy-c}

From now on we shall consider non-shift perfect fluids ($\rho \not=p$) with a non-negative pressure, $p \geq 0$. Then the energy conditions state:
\begin{equation} \label{e-c}
\hspace{-5mm} {\rm E} :  \qquad \qquad  0 \leq \pi < 1 \, , \qquad  \pi = \frac{p}{\rho} \, .
\end{equation}
Under these constraints, and for a non-dust solution, expression (\ref{pressure-parabolic-ideal}) of the pressure implies $\gamma > 1$. On the other hand, the FLRW limit $\varepsilon=0$ leads to a barotropic evolution of the form $p=(\gamma -1) \rho$. These FLRW models fulfill the energy condition (\ref{e-c}) when $\gamma < 2$, and they are the so-called $\gamma$-law models \cite{Assad-Lima}. 

Hereinafter we analyze the solutions in proposition \ref{prop-parabolic-ideal} with $1 <\gamma <  2$, which we name {\em ideal singular models}. It is worth remarking the following qualities of these solutions:
\begin{itemize}
\item[(i)] 
The metric depends on three arbitrary functions of $z$, $V_1 (z)$, $V_2 (z)$ and $W(z)$, and two effective parameters, $\kappa$ and $\gamma$. The constant $\bar{\phi}_0$ only determines an origin of time and it does not affect the metric.
\item[(ii)] 
Expression (\ref{phi-punt}) shows that the sign of the {\em amplitude parameter} $\kappa$ gives the sign of $\dot{\phi}$,  and thus it is positive in expanding models. Its square $\kappa^2$ determines the strength of the density $\rho$ and the pressure $p$. The {\em thermodynamic parameter} $\gamma$ is the only one that affects the equation of state (\ref{chi-parabolic-ideal}).
\item[(iii)] 
The solutions with $V_1 = V_2 = 0$ admit a $G_3$ on flat two-dimensional orbits. If in addition $W= constant$, then we obtain a Kantowski-Sachs model.
\item[(iv)] 
In expanding models the solutions evolve to the FLRW $\gamma$-law models.
\item[(v)] 
The solutions belong to the family of metrics considered by Szafron and Wainwright \cite{Szafron-Wain}, which was the first generalization with non-vanishing pressure of the Szekeres dust solutions.
\item[(vi)] 
The solutions also belong to the family of models proposed as two-fluid cosmologies by Lima and Tiomno \cite{Lima-Tiomno-a}. In fact, they are the parabolic subclass interpreted in a later paper \cite{Lima-Tiomno-b} as a one-component fluids furnished with a thermodynamic scheme. Here we have obtained the model by imposing a physical condition a priori: to have the hydrodynamic behavior of a generic ideal gas. Below we also obtain the full set of thermodynamic schemes and we accurately analyze their good physical behavior.
\end{itemize}

Now we achieve the analysis of the energy conditions by obtaining the spacetime domains where condition $\pi < 1$ ($\rho > p$) is fulfilled. By using (\ref{alpha-1}), the energy density (\ref{density-parabolic-ideal}) can be written in the form:
\be
\rho = \frac{3 \kappa^2 }{\phi^{3 \gamma}} 
\left[1 - \frac{2 \varepsilon (2-\gamma )}{ 3  \kappa (2-\gamma ) \phi^{\frac32 (2- \gamma)} Q + 2 \varepsilon}\right]  \, .
\ee
Then, from this expression and (\ref{pressure-parabolic-ideal}) we obtain ($\varepsilon = 1$):
\be
\rho - p = \frac{3 \kappa^2 (2-\gamma )}{\phi^{3 \gamma}} \frac{X}{X+2} \, , \qquad X \equiv 3  \kappa (2-\gamma ) \phi^{\frac32 (2- \gamma)} Q \, .
\ee
Consequently, $\rho > p$ if either $X > 0$ or $X <-2$, so that:
\begin{proposition} \label{prop-ec-parabolic-ideal}
The ideal singular models in proposition {\em \ref{prop-parabolic-ideal}} fulfill the energy conditions {\em (\ref{e-c})} in the spacetime domains where one of the following two conditions holds:
\be \label{ec-parabolic-ideal}
\kappa \, Q > 0 \, , \qquad \quad 3(2-\gamma) \kappa \, Q < -2 \phi^{-\frac32 (2- \gamma)} \, .
\ee
\end{proposition} 
Note that the first condition in (\ref{ec-parabolic-ideal}) is independent of time for both expanding ($\kappa >0$) and contracting ($\kappa <0$) models. On the other hand, for expanding models, the spatial domain where the second condition in (\ref{ec-parabolic-ideal}) holds increases with time.

%


\subsection{Compressibility conditions} 
\label{subsec-compress-c-parabolic-ideal}

The relativistic compressibility conditions  are complementary necessary requirements for physical reality of a thermodynamic perfect fluid. They were expressed \cite{Israel, Lichnero-1, Anile, Lichnero-2} by imposing constraints on the function of state $\tau = \tau(p, s)$, where $\tau = \hat{h}/n$, $\hat{h} = h/c^2$, is the so-called {\em dynamic volume}:\footnote{The {\em enthalpy index} $\hat{h}$ was introduced by Taub \cite{Taub} and named fluid index by Lichnerowicz \cite{Lichnero-1}. It is a dimensionless function of state that differs from the {\em relativistic specific enthalpy} $h$ in the constant factor $c^2$. The usual choice $c=1$ generates the identification of both functions  in literature \cite{Israel, Anile}. We prefer to use the term enthalpy index to remark its dimensionless quality that makes $\tau$ certainly be a volume.}
\begin{equation}
\hspace{-5mm} {\rm H}_1 : \qquad \qquad    (\tau'_p)_s < 0 \, , \qquad \qquad (\tau''_p)_s > 0 \, , 
 \label{cc-1}
 \end{equation}
 \begin{equation}
\hspace{-5mm} {\rm H}_2 : \qquad \qquad   (\tau'_s)_p > 0 \, , 
\label{cc-2}
\end{equation}
In \cite{CFS-CC} we have shown that the compressibility conditions H$_1$ only restrict the hydrodynamic quantities and that they can be stated in terms of the function of state $c_s^2 = \chi(\rho,p)$. For a fluid with $\chi=\chi(\pi)$, $\pi= \rho/p$, conditions (\ref{cc-1}) are equivalent to \cite{CFS-CC}:
\begin{equation}
\hspace{-5mm} {\rm H}_1 : \qquad   0 < \chi < 1 \, , \qquad   \zeta \equiv (1+\pi)(\chi-\pi) \chi'  + 2 \chi(1-\chi) > 0   \, .       \label{cc-ideal}
\end{equation}
Thus, for our models these constraints can be analyzed regardless of the functions $s(Q)$ and $r(Q)$ that define a specific thermodynamic scheme (see subsection \ref{subsec-scheme-parabolic}), and we need only to study the indicatrix function $\chi(\pi)$ given in (\ref{chi-parabolic-ideal}) in the domain $0< \pi  <1$ where the energy conditions hold. A straightforward calculation leads to:
\be \label{chi-prima}
\chi' (\pi) = \frac{2 \, \gamma \, \pi[\gamma \pi + 2(\gamma-1)]}{(\pi + 1)^2 (\pi + \gamma-1)^2} > 0 \, ,  \qquad \chi(0) =0 \, , \qquad \chi(1) = 1 \, .
\ee
Consequently, $\chi(\pi)$ is an increasing function that applies the interval $]0,1[$ to $]0,1[$, and thus the first condition in (\ref{cc-ideal}) holds. On the other hand, if we use (\ref{chi-parabolic-ideal}) and (\ref{chi-prima}) to replace $\chi$ and $\chi'$ in (\ref{cc-ideal}) we obtain:
\be
\zeta  = \frac{18 \, \gamma \, \pi^3 (1- \pi)(\gamma_0 + \gamma_1 \pi)[\gamma \pi + 2(\gamma-1)]}{ (\pi + \gamma-1)^3(\pi+1)^2} > 0 \, , 
\ee
since $\gamma_0 \equiv (3 \gamma+2) (\gamma-1) > 0$ and $\gamma_1 \equiv 5 \gamma -2 > 0$. Therefore, the second condition in (\ref{cc-ideal}) (and thus H$_1$) holds. So, we have shown:
\begin{proposition} \label{prop-cc-parabolic-ideal}
The ideal singular models in proposition {\em \ref{prop-parabolic-ideal}} fulfill the compressibility conditions ${\rm H}_1$  provided that they fulfill the energy conditions {\rm E}.
\end{proposition} 

The compressibility constraint H$_2$ depends on the full set of thermodynamic quantities \cite{CFS-CC}, that is, on the choice of $s(Q)$ and $r(Q)$, and it will be analyzed in the following sections for three specific thermodynamic schemes.


\subsection{Thermodynamic schemes: entropy, matter density and temperature} 
\label{subsec-scheme-parabolic-ideal}

Now we study the full set of thermodynamics associated with the (strict, $\varepsilon =1$) ideal singular models. We must particularize the thermodynamic schemes presented in subsection \ref{subsec-scheme-parabolic} for the solutions in proposition \ref{prop-parabolic-ideal}. Note that, from (\ref{alpha-1}) and (\ref{pressure-parabolic-ideal}), we obtain
\be \label{phi-p}
\hspace{-15mm} \phi = \phi(p) \equiv  \left[\frac{3 \kappa^2(\gamma-1)}{p}\right]^{\frac{1}{3 \gamma}}, \qquad \alpha = \alpha(p) \equiv \frac{2}{3 \kappa (2- \gamma)} \left[\frac{p}{3 \kappa^2(\gamma-1)}\right]^{\frac{2-\gamma}{2 \gamma}}  . 
\ee
Then, taking into account (\ref{phi-p}) and (\ref{density-parabolic-ideal}) we can obtain :
\be \label{Q-alpha-rho-p-1}
\hspace{-15mm}  \phi^3 ( \alpha + Q) =   \frac{K \sqrt{p}}{\rho(\gamma-1) - p} \, , \qquad K  \equiv - 2 \sigma \sqrt{(\gamma-1)/3}  \,  , \qquad \sigma \equiv \frac{\kappa}{|\kappa|}    .
\ee
Consequently, we can determine an explicit expression for the function $Q(\rho,p)$ given in (\ref{Q-rho-p}):  
\be \label{Q-rho-p-1}
\hspace{-15mm} Q = Q(\rho,p) \equiv   \tilde{K}  \, p^{\frac{2- \gamma}{2 \gamma}} \frac{\rho - p}{\rho(\gamma-1) - p} \, , \qquad \tilde{K} \equiv -\frac{2 \sigma (\gamma -1)^{\frac{3 \gamma -2}{2 \gamma}}}{\sqrt{3} (2 - \gamma) (3 \kappa^2)^{\frac{1}{\gamma}}} \, .
\ee

On the other hand, the functions $\lambda(t)$ and $\mu(t)$ defined in (\ref{lambda-mu}) can be computed in terms of $\phi$ and in terms of $p$ by using (\ref{phi-p}) and (\ref{phi-punt}):
\begin{eqnarray} \label{lambda-p}
\lambda =  \frac{4 \kappa (\gamma-1)}{2-\gamma} \phi^{-3 \gamma/2}= c_{\lambda}  \, \sqrt{p} \,  , \quad \qquad c_{\lambda} \equiv \frac{4 \sigma \sqrt{\gamma-1}}{\sqrt{3}(2-\gamma)}   \, , \\
\mu =  3 \kappa^2 \gamma \phi^{-3 (\gamma-1)}= c_{\mu} \, p^{1- \frac{1}{\gamma}}  , \quad \qquad   c_{\mu} \equiv \frac{ \gamma(3 \kappa^2)^{\frac{1}{\gamma}}}{(\gamma-1)^{1- \frac{1}{\gamma}}} > 0  \, . \label{mu-p}
\end{eqnarray}
Then, substituting all these formulas in expressions (\ref{s-n-parabolic}), (\ref{T-parabolic}) and (\ref{T-parabolic-b}) for $n$, $s$ and $\Theta$, we obtain:
\begin{proposition} \label{prop-scheme-parabolic-ideal}
The thermodynamics associated with the ideal singular models given in proposition {\em \ref{prop-parabolic-ideal}} are determined by a specific entropy $s$ and a matter density $n$ of the form:
\be  \label{s-n-parabolic-ideal}
s = s(Q) \equiv s(\rho, p) \, , \qquad  n =  \frac{\rho(\gamma-1) - p} {K \, r(Q) \, \sqrt{p}} \equiv n(\rho,p) \, .  
\ee
where $s(Q)$ and $r(Q)$ are arbitrary real functions of the function of state $Q=Q(\rho,p)$ given in {\em (\ref{Q-rho-p-1})}. Moreover the temperature is of the form:
\be  \label{T-parabolic-ideal}
\Theta = \ell(Q) \lambda(p) + m(Q) \mu(p)  \, , 
\ee
where $\lambda(p)$ and $ \mu(p)$ are given in {\em (\ref{lambda-p})} and {\em (\ref{mu-p})}, and
\be  \label{T-parabolic-b-ideal}
 \ell(Q) \equiv \frac{r'}{s'} \, , \qquad  m(Q) \equiv \frac{1}{s'}[Q r' + r]   \, .
\ee
\end{proposition}
%


\section{Models with a generic ideal gas thermodynamic scheme} 
\label{sec-idelagas}

When the hydrodynamic quantities $(u, \rho,p)$ fulfill the generic ideal gas constraint (\ref{chi-gas-ideal}), a thermodynamic scheme modeling a generic ideal gas in l.t.e. exists. Obtaining this scheme solves the {\em restricted inverse problem} for the indicatrix function $\chi = \chi(\pi)$, a problem that was analyzed in \cite{CFS-LTE}:
\begin{lemma}  \label{lemma-ideal}
If for a generic ideal gas $c_s^2 = \chi(\pi) \not= \pi$ is the square of the speed of sound then, in terms of the hydrodynamic quantities $(\rho,p)$, the specific internal energy
$\epsilon$, the temperature $\Theta$, the matter density $n$ and the specific entropy $s$ are given, respectively, by:
\begin{eqnarray}
\epsilon(\rho,p) = \epsilon(\pi) \equiv e(\pi)-1 \, , \quad \quad
\Theta(\rho,p) = \Theta(\pi) \equiv {\pi \over k} e(\pi) \, , \label{e-t-ideal} \\
n(\rho,p) = {\rho \over e(\pi)} \, ,  \qquad  \quad    \qquad  \
\quad s(\rho,p) = k \ln \frac{f(\pi)}{\rho} \, , \quad  \quad \ \  \label{n-s-ideal} 
\end{eqnarray}
the generating functions $e(\pi)$ and $\phi(\pi)$ being, respectively,
\begin{eqnarray}
e(\pi) = e_0 \exp\{\! \! \int \! \! \psi(\pi)d\pi \} \, , \quad  \qquad \psi
(\pi) \equiv \frac{\pi}{(\chi(\pi)-\pi)(\pi+1)} \,  , \label{e-pi} \\
f(\pi) = f_0 \exp\{\! \! \int \! \! \phi(\pi)d\pi\} \, , \quad \qquad
\phi(\pi) \equiv {1 \over \chi(\pi)-\pi} \, . \qquad \qquad  \
\label{f-pi}
\end{eqnarray}
\end{lemma}

For our models we must determine the generating functions $e(\pi)$ and $q(\pi)$ from the expression (\ref{chi-parabolic-ideal}) of $\chi(\pi)$, and then we obtain:
\begin{equation}
e(\pi) = e_0 \, \frac{[(\gamma-1) - \pi ]^{\frac{2(\gamma-1)}{2-\gamma}}}{(1- \pi)^{\frac{\gamma}{2-\gamma}}}    , \quad 
f(\pi) = f_0 \, \frac{1}{\pi} \left[\frac{(\gamma-1) - \pi }{1- \pi}\right]^{\frac{2 \gamma}{2-\gamma}}  . 
\label{e-h-pi-parabolic}
\end{equation}
And substituting in (\ref{e-t-ideal}) and (\ref{n-s-ideal}), we determine the expressions for $n$, $s$ and $\Theta$:
\begin{proposition}
The matter density $n$, the specific entropy $s$ and the temperature $\Theta$ of the ideal singular models with a generic ideal gas scheme take the expressions: 
\begin{eqnarray}
n(\rho,p) = \, \frac{(\rho-p)^{\frac{\gamma}{2-\gamma}}}{e_0 \, [ \rho ( \gamma-1) - p]^{\frac{2(\gamma-1)}{2-\gamma}}} \, , \qquad \Theta(\rho, p) = \frac{p}{k\, n(\rho,p)} \, , \label{n-idealgas} \\
s(\rho,p) = s_0 + k \ln \left(\frac{1}{p} \left[\frac{\rho (\gamma-1)-p}{\rho-p}\right]^{\frac{2 \gamma}{2-\gamma}}\right)  \, . 
\label{s-idealgas}
\end{eqnarray}
\end{proposition}
The above generic ideal gas scheme must correspond to a specific choice of the functions $r(Q)$ and $s(Q)$. We can determine these functions by identifying the above generic ideal gas expressions for $n$, $s$ and $\Theta$ with the generic ones given in proposition \ref{prop-scheme-parabolic-ideal}. Indeed, matching up expressions for $n$ and $s$ provided in (\ref{s-n-parabolic-ideal}) with those given in (\ref{n-idealgas}) and (\ref{s-idealgas}) we obtain:
\be \label{b-s-Q}
r(Q) = \tilde{e}_0 \, |Q|^{-\frac{\gamma}{2-\gamma}} \, , \qquad s(Q) = s_0 - \frac{2 k \gamma}{2-\gamma} \ln |Q| \, .
\ee
Note that if we use (\ref{b-s-Q}) to determine the functions $\ell(Q)$ and $m(Q)$ given in (\ref{T-parabolic-b-ideal}), then the expression for $\Theta$ given in (\ref{T-parabolic-ideal}) is coherent with that given in (\ref{n-idealgas}). 

It is worth remarking that the indicatrix function (\ref{chi-parabolic-ideal}) only approximates that of a classical ideal gas at zero-order: $\chi(0) = 0$, $\chi'(0)=0 \not= \gamma$ \cite{CFS-CIG}. Thus, our models do not show a hydrodynamic behavior similar to that of a classical ideal gas at low temperatures. 

We finish our study of the generic ideal gas scheme by analyzing the compressibility condition H$_2$. In \cite{CFS-CC} we have shown that for a generic ideal gas this thermodynamic constraint can also be stated in terms of the hydrodynamic function of state $\chi=\chi(\pi)$. More precisely, condition (\ref{cc-2}) holds if, and only if, 
\be
\hspace{-5mm} {\rm H}_2 : \qquad \qquad    \xi \equiv (2 \pi + 1) \chi(\pi) - \pi > 0 \, .
\ee
For the indicatrix function $\chi(\pi)$ given in (\ref{chi-parabolic-ideal}), we obtain for $\xi$:
\be
\xi(\pi) = \frac{\chi}{2 \gamma \pi}[(4 \gamma -1) \pi^2 + \gamma \pi - (\gamma-1)] \, .
\ee
A straightforward calculation shows (if $\gamma>1$) that $\xi(\pi)$ vanishes in the interval $]0,1[$ for the value:
\be \label{pi-m}
\pi_m = \frac{1}{2(4 \gamma -1)}[\sqrt{17 \gamma^2 - 20 \gamma+4} - \gamma]   \, ,
\ee
and it is positive for $\pi > \pi_m$. Consequently, we can state.
\begin{proposition} \label{propo-cc2-ideal}
The ideal singular models with a generic ideal gas thermodynamic scheme fulfill the compressibility condition ${\rm H}_2$ in the domain where $p/\rho = \pi \in ]\pi_m, 1[$, with $\pi_m$ depending on the thermodynamic parameter $\gamma$ as {\em (\ref{pi-m})}.
\end{proposition}
%


\section{The Lima-Tiomno models} 
\label{sec-LT}

The generic ideal gas thermodynamic scheme presented in the above section is just one of the possible thermodynamics that can be associated with each of the ideal singular solutions. As stated in proposition \ref{prop-scheme-parabolic-ideal}, these solutions model the evolution in l.t.e. of a wide family of perfect fluids defined by each election of the two functions $s(Q)$ and $r(Q)$. Next we will see that some of these elections also model inviscid fluids with a non-vanishing conductivity coefficient.

In the framework of a macroscopic theory for non-perfect fluids, the fluid velocity is submitted to severe constraints when the dissipative fluxes (anisotropic pressures, bulk viscous pressure and energy flux) vanish \cite{ReZa}. Indeed, if none of the thermal transport coefficients (shear viscosity, bulk viscosity and thermal conductivity) is zero, the macroscopic constitutive equations imply that the fluid flux defines a time-like Killing vector.  Nevertheless, for inviscid fluids (vanishing shear and bulk viscosity coefficients), thermal equilibrium only implies that the fluid acceleration is constrained by the relation:\footnote{This equation generalizes the well-known result by Tolman \cite{Tolman-30, Tolman} on selfgravitating spheres in thermal equilibrium, and can be obtained from the relativistic Fourier law by Ekcart \cite{Eckart}, and also from the Fourier laws proposed in causal extended thermodynamics \cite{Israel-b, Israel-St, Jou-Casas, ReZa}.}
\be \label{Fourier}
a = - \perp d \ln \Theta \, ,
\ee
where $a$ is the fluid acceleration and $\perp$ denotes the orthogonal projection to the fluid velocity.  
 
In our models we have a geodesic motion, $a=0$. Thus, (\ref{Fourier}) implies a homogeneous temperature, $\Theta = \Theta(t)$ or, equivalently, it must be a function of $p$. The temperature (\ref{e-t-ideal}) depends on $\pi$ and then on $\rho$, and thus the model with a generic ideal gas scheme considered in previoius section is only compatible with vanishing conductivity.

In this section and in the following one, we consider two thermodynamic schemes that are compatible with non-vanishing conductivity. Note that from (\ref{T-parabolic-ideal}) the demand $\Theta = \Theta(p)$ is only consistent with the thermodynamic schemes that fulfill the conditions $\ell'(Q) = m'(Q)=0$. 

The first scheme that we examine defines a model already considered by Lima and Tiomno \cite{Lima-Tiomno-b} and that we review from our approach. If we select $\ell(Q) = \ell_0 \not=0$, and $m(Q)=0$, then from (\ref{T-parabolic-b-ideal}) we obtain:
\be
r(Q) = \frac{r_0}{Q} \, , \qquad s(Q) = s_0 + \frac{r_0}{\ell_0 Q} \, .
\ee
Then, if we take the arbitrary constants $r_0$ and $\ell_0$ such that $\kappa \, \ell_0 > 0$ and $r_0 > 0$, we  obtain:
\begin{proposition} \label{propo-Lima}
The ideal singular models admit thermodynamics compatible with non-vanishing conductivity defined by a matter density $n$, a temperature $\Theta$ and an entropy $s$ given by:
\begin{eqnarray}
n(\rho,p) = n_1 (\rho-p) p^{\frac{1}{\gamma}-1}  \, , \qquad \Theta(\rho, p) = \Theta_1 \sqrt{p} \, , \label{n-Lima} \\
s(\rho,p) = s_0 - s_1  \frac{\rho (\gamma-1)-p}{\rho-p} \, p^{- \frac{2-\gamma}{2 \gamma}}  , \qquad s_1 \equiv  \frac{2}{n_1 \Theta_1 (2-\gamma)} > 0 \, . 
\label{s-Lima}
\end{eqnarray}
where $n_1$ and $\Theta_1$ are arbitrary positive constants.
\end{proposition}
We have that $\Theta > 0$, and $n >0$ if $\rho>p$. Then, the expression (\ref{s-Lima}) for $s$ is well defined, and from (\ref{pressure-parabolic-ideal}), the formulas (\ref{n-Lima}) are equivalent to:
\be \label{rho-p-n-lima}
\rho = p + \frac{1}{n_1} n\, p^{1-1/ \gamma} \, , \qquad \Theta \propto \phi^{- 3 \gamma/2} \, .
\ee
These two expressions can be found in \cite{Lima-Tiomno-b}, and they show that, indeed, the thermodynamic scheme given in proposition \ref{propo-Lima} is the one considered by Lima and Tiomno. 

Now we study the compressibility condition ${\rm H}_2$ for the Lima-Tiomno thermodynamic scheme. In \cite{CFS-CC} we have shown that condition ${\rm H}_2$ is equivalent to:
\be \label{H2-Theta}
\hspace{-5mm} {\rm H}_2 : \qquad \qquad  2 n \Theta > \frac{1}{s_{\rho}'}    \, .
\ee
From (\ref{s-Lima}), for the Lima-Tiomno models we have:
\be
s'_{\rho} = - s_1 \frac{(2-\gamma)p}{(\rho-p)^2} \, p^{\frac{3 \gamma-2}{2 \gamma}} < 0   \, .
\ee
Thus, condition (\ref{H2-Theta}) holds and, consequently, we can state:
\begin{proposition} \label{propo-cc2-Lima}
The Lima-Tiomno schemes given in proposition {\em \ref{propo-Lima}} fulfill the compressibility condition ${\rm H}_2$, and the associated matter density $n$ and temperature $\Theta$ are positive in the domain where the energy conditions {\rm E} hold.
\end{proposition}
%


\section{Models with the FLRW-limit temperature} 
\label{sec-T-FLRW}

It is worth remarking that the Lima-Tiomno models presented in the section above have a temperature $\Theta$ that is a power of the metric function $\phi$ (see (\ref{rho-p-n-lima})). But the exponent $-3 \gamma/2$ does not coincide with the exponent of the power expression for the temperature in the $\gamma$-law model of the FLRW limit. In this case we have $\Theta \propto \phi^{- 3 (\gamma-1)}$ \cite{Assad-Lima}. This fact was pointed out by Lima and Tiomno \cite{Lima-Tiomno-b} and they comment on this unexpected result. The Lima-Tiomno models were later analyzed in \cite{QS-1995}, and the incompatibility between the thermodynamic scheme by Lima-Tiomno and the FLRW limit was again remarked.

Our approach enables us to shed light on this situation. Indeed, proposition \ref{prop-scheme-parabolic-ideal} shows that the singular models admit a wide family of thermodynamic schemes with homogeneous temperature. They are determined by two functions $s(Q)$ and $r(Q)$ submitted to constraints (\ref{T-parabolic-b-ideal}), with $\ell(Q) = \ell_0$ and $m(Q)=m_0$, $\ell_0$ and $m_0$ being two arbitrary constants. Then, (\ref{T-parabolic-ideal}), (\ref{lambda-p}) and (\ref{mu-p}) show that the temperature $\Theta$ is the sum of two power functions of the pressure, and then of the metric function $\phi$. We have selected in the above section $\ell_0 \not=0$ and $m_0 = 0$ and we have obtained the power function of the Lima-Tiomno model. Now we show that the other possibility of acquiring a single power function ($\ell_0 =0$ and $m_0 \not = 0$) leads to a model with the temperature of the FLRW limit. 

Hence, now we select $\ell_0 =0$ and $m_0 \not = 0$. Then, from (\ref{T-parabolic-b-ideal}) we obtain:
\be
r(Q) = r_0 \, , \qquad s(Q) = s_0 + \frac{r_0}{m_0} Q  \, .
\ee
Then, if we take the arbitrary constants $r_0$ and $m_0$ such that $m_0 > 0$ and $\kappa \, r_0 < 0$, we  obtain:
\begin{proposition} \label{propo-FLRW}
The ideal singular models admit thermodynamics with the same temperature as in the $\gamma$-law models of the FLRW limit. The matter density $n$, the temperature $\Theta$ and the entropy $s$ are given by:
\begin{eqnarray}
n(\rho,p) = \frac{n_1}{\sqrt{p}} [\rho (\gamma-1)-p] \, , \qquad \Theta(\rho, p) = \Theta_1\, p^{1- \frac{1}{\gamma}}  \, , \label{n-FLRW} \\
s(\rho,p) = s_0 + s_1  \frac{\rho-p}{\rho (\gamma-1)-p} \, p^{ \frac{2-\gamma}{2 \gamma}}  , \qquad s_1 \equiv  \frac{\gamma}{n_1 \Theta_1} > 0   \, . 
\label{s-FLRW}
\end{eqnarray}
where $n_1$ and $\Theta_1$ are arbitrary positive constants.
\end{proposition}
We have that $\Theta > 0$, and $n >0$ if $(\gamma -1) \rho>p$. Then, expression (\ref{s-FLRW}) for $s$ is well defined, and the energy conditions hold. Moreover, from (\ref{pressure-parabolic-ideal}), expressions (\ref{n-FLRW}) can be written as:
\be \label{rhp-p-n-lima}
\rho = \frac{1}{\gamma-1} (p + \frac{1}{n_1} n \, \sqrt{p}) \, , \qquad \Theta \propto \phi^{-3(\gamma-1)} \, ,
\ee
and we see that, indeed, we obtain the temperature of the $\gamma$-law models of the FLRW limit.

The spacetime domain where condition $(\gamma -1) \rho>p$ holds can be obtained by using expressions (\ref{alpha-1}), (\ref{pressure-parabolic-ideal}) and (\ref{density-parabolic-ideal}) for $\alpha$, $p$ and $\rho$. Then, a straightforward calculation leads to the second constraint in (\ref{ec-parabolic-ideal}). 

Finally we study the compressibility condition ${\rm H}_2$ for this thermodynamic scheme. From (\ref{s-FLRW}) we obtain:
\be
s'_{\rho} = - s_1 \frac{(2-\gamma)p^{\frac{2+\gamma}{2 \gamma}}}{[\rho (\gamma-1)-p]^2}   < 0  \,   .
\ee
Thus, condition (\ref{H2-Theta}) holds and, consequently, we can state:
\begin{proposition} \label{propo-cc2-FLRW}
The ideal singular models with the thermodynamic scheme given in proposition {\em \ref{propo-FLRW}} fulfill the compressibility condition ${\rm H}_2$, and the associated matter density $n$ and temperature $\Theta$ are positive and the energy conditions hold in the spacetime domain defined by the constraint
%
%
\be \label{ec-FLRW}
3(2-\gamma) \kappa \, Q < - 2 \phi^{-\frac32 (2- \gamma)} \, .
\ee%
\end{proposition}
Note that the spatial domain where (\ref{ec-FLRW}) holds increases with time for the expanding models.


\section{Summary of the results} 
\label{sec-summary}

An important task in Relativity is the study of the physical meaning of the formal perfect fluid solutions to the field Einstein equations. At present, a wide family of such solutions is known without specific physical meaning. The energy conditions \cite{Plebanski} are necessary conditions for physical reality. But complementary physical requirements must be imposed on the hydrodynamic quantities $\{u, \rho, p\}$ if we look for solutions that model thermodynamic perfect fluids in local thermal equilibrium. Our hydrodynamical approach to the macroscopic l.t.e. \cite{CFS-LTE} provides a tool to impose these requirements by means of the hydrodynamic sonic condition (\ref{lte-chi}). Moreover, it allows us to analyze the necessary macroscopic physical constraints of this family of solutions, and to solve the inverse problem for obtaining the specific thermodynamic interpretation. Additionally, the positivity of some thermodynamic quantities and the compressibility conditions \cite{Israel, Lichnero-1} must be also imposed to obtain a coherent shock theory. Our hydrodynamic approach to the compressibility conditions \cite{CFS-CC} furnishes a tool to set this physical requirement.

In this work we have analyzed the hydrodynamic sonic condition (\ref{lte-chi}) for the class II Szekeres-Szafron metrics, and we have shown that three families of thermodynamic solutions in l.t.e. can be distinguished: the metrics admitting a G$_3$ on S$_2$, the singular models, and the regular models (proposition \ref{prop-II}). Here we restrict ourselves to the study of the singular models, and we offer the canonical form for the metric line element and the expression of the energy density $\rho$ and pressure $p$ (subsection \ref{subsec-metric-parabolic}). We also determine the compatible thermodynamics by obtaining the general expressions for the specific entropy $s$, the matter density $n$ and the temperature $\Theta$ (propositions \ref{prop-s-n-parabolic} and \ref{prop-T-parabolic}), as well as an implicit expression for the square of the speed of sound $\chi(\rho,p)$ (proposition \ref{prop-chi-parabolic}).

We have studied in depth the ideal singular models, which are defined by the generic ideal gas hydrodynamic sonic condition, $\dif \chi \wedge \dif \pi = 0$, $\pi = p/\rho$, $\chi = \dot{p}/\dot{\rho}$. We have integrated this equation and we have obtained the expression of the metric functions, of the hydrodynamic quantities, pressure $p$ and energy density $\rho$, and of the square of the speed of sound $c_s^2 = \chi(\pi)$. These expressions have allowed us to study the space-time domain where the ideal gas singular models fulfill the energy conditions E and the compressibility conditions H$_1$. Now, the compatible thermodynamics are explicitly given up to two functions of the entropy. Table \ref{table-1} summarize all these results.

\begin{table}[h]

\begin{tabular}{cl}
\hline \\[-5mm] \hline

 & \qquad  \quad Ideal singular models: $\displaystyle \chi = \chi(\pi) \neq \pi$ \phantom{\Large $\frac{A}{B}$}
 \\[2mm] \hline
 Metric line element &   \quad  \qquad $ \displaystyle \dif s^2 = - \dif t^2 + \phi^2 \Big[
 (\epsilon \alpha + Q )^2 \dif z^2 + \dif x^2 + \dif y^2 \Big] $ \phantom{\LARGE $\frac{A}{B}$}
 \\[2mm]
  \hline
  &   \qquad  \quad   $\displaystyle Q= V_1 (z) x + V_2 (z) y + 2 W(z) $    \phantom{\Large $\frac{A}{B}$} \\
Metric functions    &  \\[-5mm]
   & \qquad  \quad   $\displaystyle \phi(t) =  \Big[ \frac{3}{2} \kappa t + \bar{\phi}_0 \Big]^\frac{2}{3 \gamma} $,   \qquad   $\displaystyle \alpha(t) = \frac{2}{3 \kappa (2-\gamma)} \, \phi^{\frac{3}{2}
   (\gamma -2) } $ 
  \\[3mm] \hline
 $p$ & \qquad \quad  $\displaystyle  p(t)  = \frac{3 \kappa^2 (\gamma -1)}{\phi^{3 \gamma}} $  \phantom{\LARGE $(\frac{A}{B})^{\Big(C \Big)}$}
 \\[3mm]
 \hline
 $\rho$  & \quad \qquad    $ \displaystyle \rho(t, Q) =\frac{3 \kappa^2}{\phi^{3 \gamma}} - \frac{ 2 \epsilon
 \kappa}{ (\epsilon \alpha + Q) \ \phi^{3 (1 + \frac{\gamma}{2})} }
 $  \phantom{\LARGE $(\frac{A}{B})^{\Big(C \Big)}$}  \\[4mm] \hline
 $\displaystyle {c_s}^2 $ &  \qquad \quad $ \displaystyle {c_s}^2  = \chi(\pi) = \frac{2 \gamma \pi^2 }{(\pi + 1) (\pi + \gamma -
 1)}$, \  \  \ $\displaystyle \pi= \frac{p}{\rho}$ \phantom{\LARGE $(\frac{A}{B})^{\Big(C \Big)}$}  \hspace{-1mm} \\[3.2mm] \hline
 E &  \qquad \quad   $\displaystyle \kappa Q >0$ \qquad or \qquad $3 (2 - \gamma ) \kappa Q < -2 \phi^{-
 \frac{3}{2} (2 - \gamma)} $ \phantom{\Large $(\frac{A}{B})^{\Big(C \Big)}$} \hspace{-3mm}  \\[1mm] \hline
H$_1$  & \qquad   \quad  E    \phantom{\Large $(\frac{A}{B})$}\\[1mm] \hline
$n$ &
\qquad \quad  $ \displaystyle n(\rho, p)= \frac{\rho(\gamma-1) - p} {K \,
r(Q) \, \sqrt{p}} $  \phantom{\Large $(\frac{A}{B})^{\Big(C \Big)}$}
\\[3.8mm] \hline
$\Theta$ & \quad  \qquad 
$\displaystyle \Theta(\rho, p)  = \ell(Q) \lambda(p) + m(Q) \mu(p) $ \phantom{\Large $(\frac{A}{B})$}
\\[1mm]
\hline
$s$ & \qquad \quad 
$s(\rho, p) = s(Q)$  \phantom{\Large $(\frac{A}{B})$}  \\[1mm] \hline \\[-5mm] \hline
\end{tabular}
\caption{This table shows the metric tensor and the hydrodynamic quantities, $p$, $\rho$ and $\chi(\pi)$, of the ideal singular models. We have two effective parameters: the thermodynamic parameter $\gamma$, which ranges in the interval $]1,2[$, and the amplitude parameter $\kappa \not=0$, $\kappa>0$ for expanding models and $\kappa<0$ for contracting models. It also provides the hydrodynamic constraints imposed by the energy conditions E and the compressibility condition H$_1$. The generic compatible thermodynamics are presented in the last three rows: $r(Q)$ and $s(Q)$ are two arbitrary functions of the quantity $Q(\rho,p)$ given in (\ref{Q-rho-p}), and $\ell(Q)$, $m(Q)$, and $\lambda(p)$, $\mu(p)$ are given in (\ref{T-parabolic-b-ideal}) and (\ref{lambda-p}-\ref{mu-p}).}
\label{table-1}
\end{table}

\begin{figure}
\centerline{
\parbox[c]{0.50\textwidth}{\includegraphics[width=0.48\textwidth]{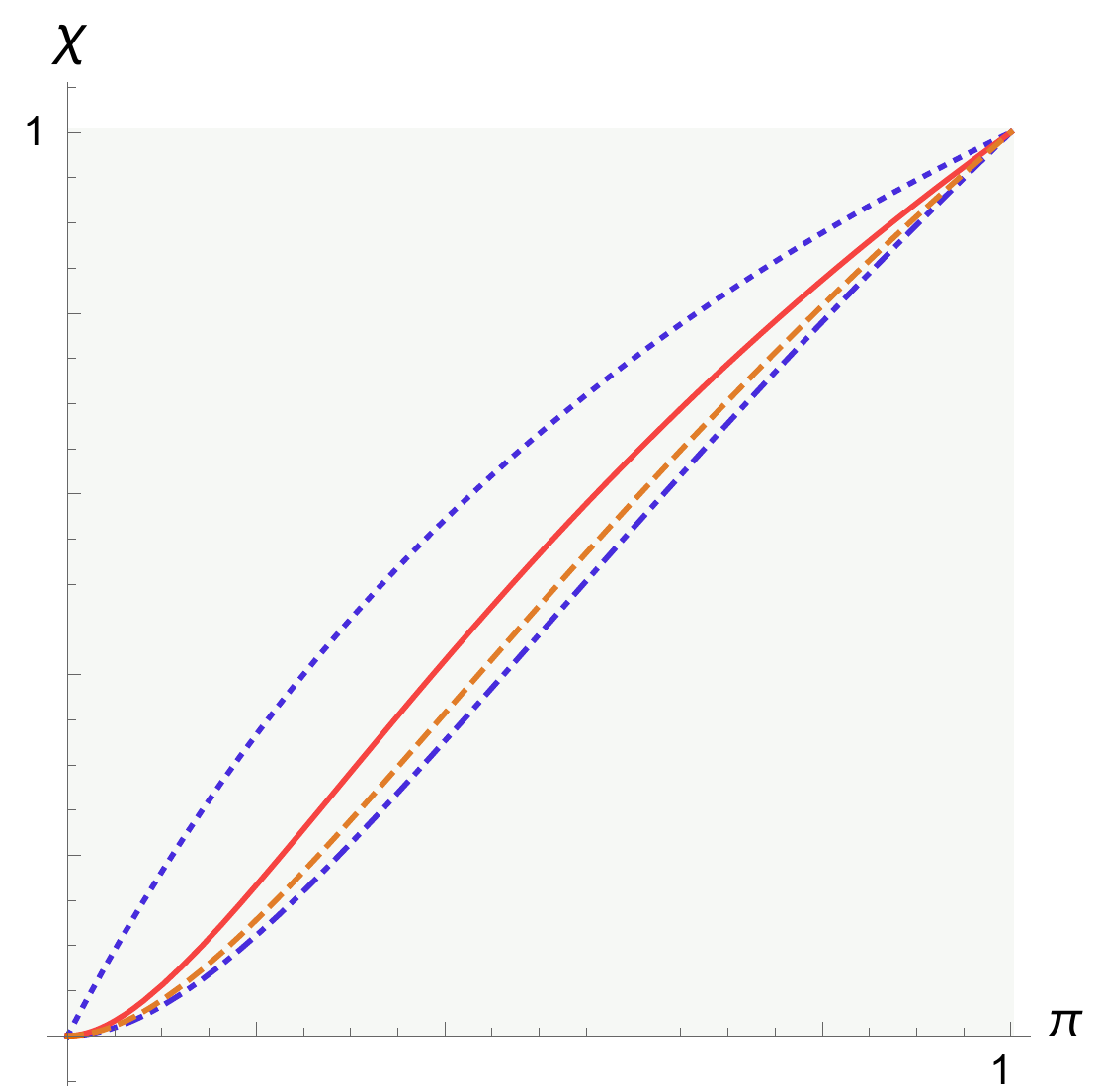}}
\parbox[c]{0.50\textwidth}{\includegraphics[width=0.48\textwidth]{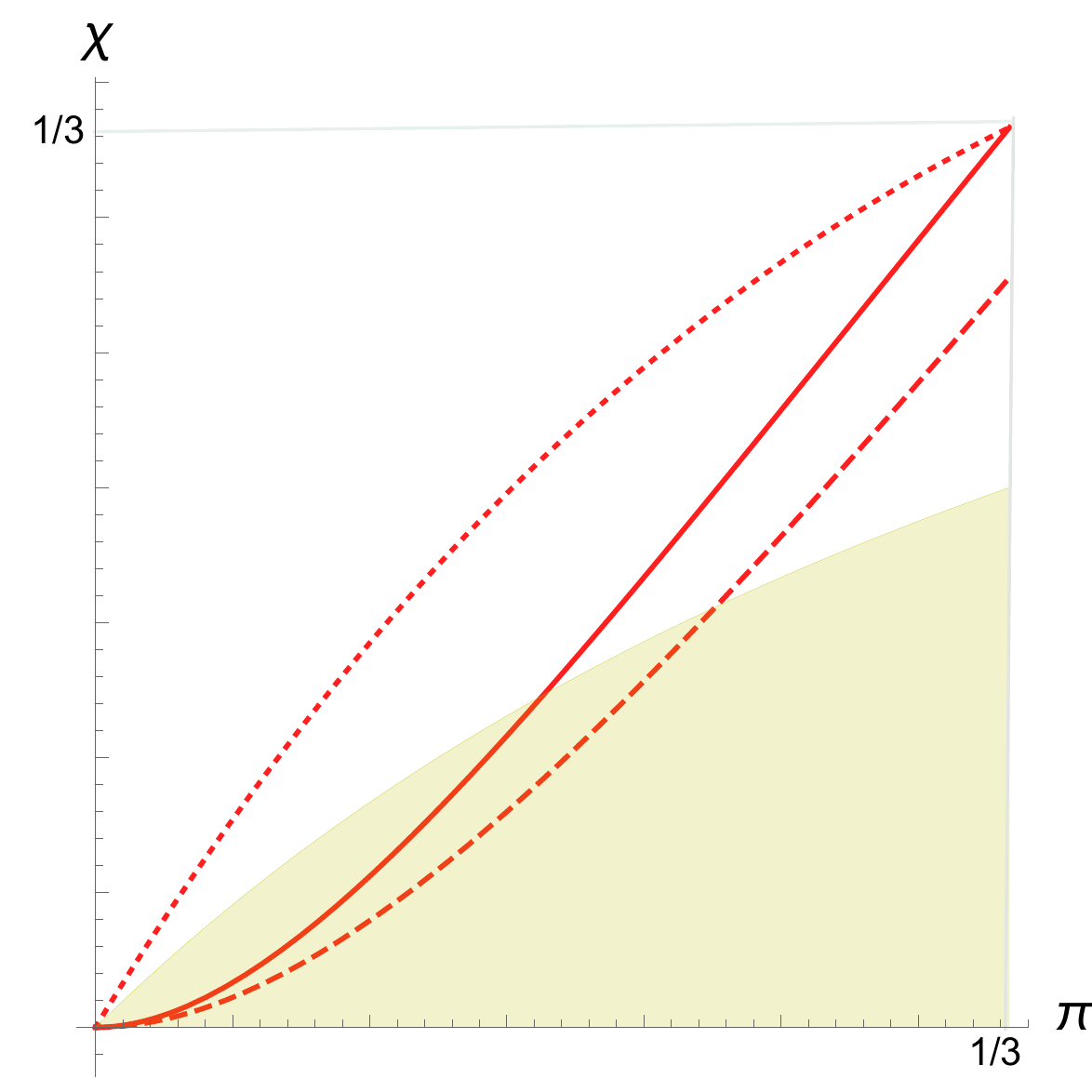}}}
\caption{On the left, the graph of the function $\chi= \chi(\pi)$ that gives the square of the speed of sound $c_s^2 = \chi$ as a function of the hydrodynamic quantity $\pi=p/\rho$ for $\gamma = 5/3$ (red dashed line) and $\gamma=4/3$ (red solid line). For all the values of $\gamma$ the graphs are between the limiting cases $\gamma=1$ (blue dotted line, at the top) and $\gamma=2$ (blue 
dotdashed line, at the botton), and the compressibility condition H$_1$ is satisfied. On the right, the graph of the same cases $\gamma = 5/3$ and $\gamma = 4/3$ in the interval $[0/1/3]$, as well as the graph of $c_s^2$ for the relativistic Synge gas (red dotted line). For the generic ideal gas thermodynamic scheme the compressibility condition H$_2$ also constraints the indicatrix function $\chi(\pi)$: the shaded area in the plot is forbidden. Note that then the hydrodynamic variable $\pi$ is limited to an interval $]\pi_m, 1[$, and the case $\gamma = 4/3$ approaches the Synge gas at high temperatures. This constraint on the quantity $\pi$ does not apply to the  thermodynamic schemes considered in sections 6 and 7.
\label{Fig-1}}
\end{figure}

We know \cite{CFS-CC} that the compressibility conditions H$_1$ impose constraints on the square of the speed of sound, which for the ideal singular models is the function $c_s^2 = \chi(\pi)$ given in the sixth row of table \ref{table-1}. The left plot in figure \ref{Fig-1} shows the graph of this function for different values of $\gamma$.

Among all admissible thermodynamics, we have extended our analysis to three of them. The first one defines a generic ideal gas and it has inhomogeneous temperature. The other two thermodynamics have homogeneous temperature, and so they are compatible with the current relativistic heat equations. This means that these two thermodynamic schemes model either perfect fluids in l.t.e. or inviscid non-perfect (with non-vanishing conductivity coefficient) fluids in thermal equilibrium.
The space-time domains where the thermodynamic compressibility condition H$_2$ holds have been obtained. The right plot in figure \ref{Fig-1} shows the region where the compressibility condition H$_2$ hold for the generic ideal gas scheme presented in section 5. This constraint on the hydrodynamic quantity $\pi$ does not apply for the other two considered thermodynamic schemes. Table 2 summarizes all these results.

\begin{table}[h]
\begin{tabular}{cllll}
\hline \\[-5mm] \hline
  &   $n(\rho, p) $  & \  $\Theta(\rho, p)$  & \ \ $s(\rho, p) $ & \ \ H$_2$  
  \hspace{-10mm} \phantom{\Large $(\frac{A}{B})$} 
  \\[1mm]
  \hline
\hspace{-3mm}   IG  &   $\displaystyle \frac{{e_0}^{-1}
(\rho \!- \!p)^{\frac{\gamma}{2-\gamma}}}{ \left[ \rho(\gamma  \!- \!1) \!- \! p
\right]^\frac{2(\gamma -1)}{2-\gamma}} $ &  \  $\displaystyle \frac{p}{k
n(\rho, p) } $  & \ $ \ s_0 \! + \! k  ln \! \left(\!\! \frac{1}{p}\! \! \left[\frac{\rho
(\gamma-1)-p}{\rho-p}\right]^{\!\frac{2 \gamma}{2-\gamma}} \! \!\right) $  &
\ \ $\pi \in ] \! -  \!  \pi_m , 1 [ $   
\hspace{-20mm}  \phantom{\LARGE $(\frac{A}{B})^{\Big(C \Big)}$}  
\\[5mm] \hline
\hspace{-3mm}  LT &  $ \displaystyle n_1 (\rho - p) \, p^{\frac{1}{\gamma}-1} $ & \
 $\displaystyle \Theta_1  \sqrt{p} $ & \ \ $\displaystyle  s_0 \! - \! s_1  \frac{ \rho(\gamma  \!- \!1) \!- \! p}{\rho-p}  p^{
\frac{-(2-\gamma)}{2 \gamma}} $ & \ \ E 
\hspace{-20mm} \phantom{\Huge $(\frac{A}{B})$}  
\\[2.8mm] \hline
\hspace{-2mm}FLRW & $\displaystyle  n_1 \frac{ \rho(\gamma  \!- \!1) \!- \! p}{\sqrt{p}} $ & \
 $ \displaystyle   \Theta_1 \,
  p^{1 - \frac{1}{\gamma}} $  & \ \ $ \displaystyle s_0 \! + \!s_1 \frac{\rho-p}{\rho(\gamma  \!- \!1) \!- \! p}  p^{
\frac{2-\gamma}{2 \gamma}}$ & \  \ $\displaystyle \kappa  Q \! < \!  \! \frac{-2
\phi^{-\frac32 (2- \gamma)}}{3(2-\gamma)} $  
\hspace{-16.5mm} \phantom{\Huge $(\frac{A}{b})$} \\[3.2mm] 
\hline \\[-5mm] \hline
\end{tabular}
\caption{This table provides the explicit expression of the matter density $n$, the temperature $\Theta$ and the specific entropy $s$ in terms of the hydrodynamic quantities $\rho$ and $p$ for three specific thermodynamics, the generic ideal gas thermodynamic scheme (IG), the Lima-Tiomno model (LT), and the model with the temperature of the FLRW-limit (FLRW). For the three cases, the results summarized in table \ref{table-1} apply. Here, the last column shows the  constraints imposed by the thermodynamic compressibility condition H$_2$.}
\label{table-2}
\end{table}





\ack This work has been supported by the Spanish ``Ministerio de
Econom\'{\i}a y Competitividad", MINECO-FEDER project FIS2015-64552-P.


\end{document}